\documentclass[aps,superscriptaddress,showpacs,pra,10pt,twocolumn]{revtex4-1}
\usepackage{amsmath}
\usepackage{graphicx}
\usepackage{color}
\usepackage{comment}
\usepackage{multirow}
\usepackage{array}
\usepackage{longtable}
\usepackage{adjustbox}
\usepackage{mathtools} 
\usepackage{bbold}
\usepackage{float}
\usepackage{placeins}

\newcommand{\bra}[1]{\langle #1 |}
\newcommand{\ket}[1]{| #1 \rangle}

\newcommand {\be}{\begin{equation}}
\newcommand {\ee}{\end{equation}}

\newcommand\D{{\text{c}}}

\newcommand{\ba}{\begin{eqnarray}}
\newcommand{\ea}{\end{eqnarray}}

\newcommand{\ignore}[1]{}

\newcommand{\Tr}{{\mathrm{Tr}}}
\newcommand{\e}{{{e}}}

\newcommand{\beq}{\begin{equation}}
\newcommand{\eeq}{\end{equation}}
\newcommand{\beqnn}{\begin{equation*}}
\newcommand{\eeqnn}{\end{equation*}}
\newcommand{\bea}{\begin{eqnarray}}
\newcommand{\eea}{\end{eqnarray}}
\newcommand{\beann}{\begin{eqnarray*}}
\newcommand{\eeann}{\end{eqnarray*}}
\newcommand{\bes} {\begin{subequations}}
\newcommand{\ees} {\end{subequations}}

\begin{document}
%%%%%%%%%%%%%%%%%%%%%%%%%%%%%%%%%%%%%%%%%%%%%%%
\title{Stoquastic simulations of non-stoquastic superconducting flux circuits}
\author{Harel Kol-Namer}
\affiliation{Raymond and Beverly Sackler School of Physics and Astronomy, Tel Aviv University, Tel Aviv 6997801, Israel}
\author{Tom Halverson}
%\email{halverso@isi.edu}
\affiliation{Information Sciences Institute, University of Southern California, Marina del Rey, California 90292, USA}
\author{Lalit Gupta}
\affiliation{Department of Physics and Astronomy and Center for Quantum Information Science \& Technology, University of Southern California, Los Angeles, California 90089, USA}
\author{Moshe Goldstein}
\affiliation{Raymond and Beverly Sackler School of Physics and Astronomy, Tel Aviv University, Tel Aviv 6997801, Israel}
%\author{David G. Ferguson}
%\affiliation{Northrop Grumman Corporation, Linthicum, Maryland 21090, USA}
\author{Itay Hen}
\affiliation{Information Sciences Institute, University of Southern California, Marina del Rey, California 90292, USA}
\affiliation{Department of Physics and Astronomy and Center for Quantum Information Science \& Technology, University of Southern California, Los Angeles, California 90089, USA}

%\date{\today}
%%%%%%%%%%%%%%%%%%%%%%%%%%%%%%%%%%%%%%%%%%%%%%%
%            Paper Abreviations
% Adiabatic Quantum Computing - AQC
% non-stoquastic - NStoq
% stoquastic - SQ
% Quantum Monte Carlo - QMC
% radio frequency superconducting quantum interference devices  - rf-SQUID
% first order finite difference - FD
% Number of qubits - N
%%%%%%%%%%%%%%%%%%%%%%%%%%%%%%%%%%%%%%%%%%%%%%%
%---------------------------------------------------------------------------------------
   \begin{abstract}
\noindent 
There is a tremendous interest in fabricating superconducting 
flux circuits that are nonstoquastic---i.e., have positive off-diagonal matrix elements---in their qubit representation, as these circuits are thought to be unsimulable by classical
approaches due to the presence of a sign problem and thus could play a key role in the demonstration of speedups in quantum annealing protocols. We show, however, that the elimination of the sign problem in these systems is possible by the direct simulation of the flux circuits. Our approach not only obviates the reduction of flux circuits to their qubit representation but also produces results that are more in the
spirit of the experimental setup. We discuss the implications of our work, arguing that our findings cast doubt on the conception that superconducting flux
circuits represent the correct avenue for universal adiabatic quantum computers.
   \end{abstract}
%---------------------------------------------------------------------------------------
\maketitle
%%%%%%%%%%%%%%%%%%%%%%%%%%%%%%%%%%%%%%%%%%%%%%%%%%%%%%%%%%%%%%%%%%%%%%%%%%%%%%%%%%%%%%%%%%%%%%%%%%%%%
%%%%%%%%%%%%%%%%%%%%%%%%%%%%%%%%%%%%%%%%%%%%%%%%%%%%%%%%%%%%%%%%%%%%%%%%%%%%%%%%%%%%%%%%%%%%%%%%%%%%%
%---------------------------------------------------------------------------------------
\section{Introduction}
Adiabatic Quantum Computing (AQC)~\cite{RevModPhys.90.015002} is a paradigm in which computation is accomplished by evolving a quantum Hamiltonian slowly enough so as to ensure the system remains in its instantaneous ground state.  AQC is considered a leading candidate for the future of scalable robust quantum devices. It has long been known that for AQC to be universal~\cite{Biamonte:07,aharonov_adiabatic_2007}, the Hamiltonians of interest must include non-stoquastic (NStoq) terms~\cite{bravyi2009complexity,marvian:2018,Klassen2019twolocalqubit,doi:10.1137/19M1287511}, i.e., terms whose matrix representations include positive off-diagonal elements that cannot be easily transformed away~\cite{gupta2019,crosson2020designing}. In addition, NStoq Hamiltonians are not known to be efficiently simulated with classical computational methods such as Quantum Monte Carlo (QMC), owing to the relationship between non-stoquasticity and the QMC sign problem~\cite{gupta2019,marvian:2018,Wiese-PRL-05} (the reader is also referred to Refs.~\cite{Hastings2021powerofadiabatic,10.1145/3406325.3451060} for examples where AQC is advantageous over classical methods even for stoquastic systems). NStoq terms are, therefore, seen as one of the missing ingredients for the demonstration of adiabatic quantum speedups~\cite{nonStoq3,nonStoq2,nonStoq1,PhysRevB.95.184416,Alb2019,PhysRevA.95.042321,Dur2019}.

%It is due to the above that considerable effort has been put toward the practical realization of NStoq Hamiltonian and what effort that has been put forward has shown some promise, though. 
Currently, several groups are utilizing radio frequency superconducting quantum interference devices (rf-SQUIDs) to engineer NStoq systems,~\cite{Kjaergaard2020}, and preliminary success has already been reported~\cite{dwaveNonStoq}. These types of devices represent some of the current cutting edge in quantum computing technology. However, as we will discuss here, much of this effort may be in vain. While there is no argument that truly NStoq experimental qubit systems would constitute a huge step towards AQC universality and perhaps even demonstrate some form of quantum supremacy, it is important to reckon with the fact that flux qubit systems are approximations to superconducting circuits, which are not genuinely two-level systems.
%, and it is well known that these qubit approximation are not without flaw. Issues such as the leakage between the lower-level qubit states and higher eigenenergies through thermal excitation resulting is computational noise could be cause for concern as to whether or not they result in an accurate interpretation of the device physics. Just because the qubit Hamiltonian exhibits a given property, does not necessarily mean the system itself has that property.  
%In fact, we show here that superconducting flux circuits that are NStoq --- and hence unsimulable --- in their qubit representation, are in fact Stoq, or can be made Stoq, when the entire circuit is considered. This observation in turn allows us to efficiently simulate these systems, calling into question the entire approach of fabricating NStoq quantum annealers with rf-SQUID technology.
In fact, we show here that the current paradigm of flux qubit systems that are considered to be NStoq (and hence unsimulable), are Stoq, or at least can be made Stoq, when the circuit is considered. This observation in turn allows us to avoid the sign problem in these systems, calling into question the entire approach of fabricating NStoq quantum annealers with rf-SQUID technology.

%It should be noted that we are not attempting to wholly disprove the validity of the qubit approximation. 
%These models are the linchpin of most of modern quantum computing, but it is important to ascertain whether the approximation has any bearing on the actual non-stoquasticity of the circuit, and in turn the non-simulability of the device, as a whole.

The elimination of the sign problem is possible because superconducting circuits are not described by discrete spin states, but rather by conjugate variable pairs in a continuous Hilbert space. These phase-space coordinate pairs are usually expressed as flux and charge variables---which represent position- and momentum-like coordinates, respectively. 
This observation was initially pointed out by Bravyi et al.~\cite{bravyi2006complexity} who asserted that if we treat the quantum mechanics of Josephson-junction qubit systems of the ``flux-type as a collection of distinguishable (rather than bosonic or fermionic) particles the kinetic energy is a Laplacian in the continuous limit (see also Ref.~\cite{PhysRevB.69.064503}).  
Devising a combination of rotation of the charge and discretization
of flux, we explicitly show that the circuit Hamiltonian can be cast into a Stoq Hamiltonian, and so sign-problem free QMC techniques may be applied~\footnote{An independent study making similar assertions that has very recently been posted online~\cite{Ciani2021}. However, the focus and methods by which the resolution of the off-diagonal matrix elements in the kinetic energy matrix is achieved are quite different.}.

It is important to note that while our approach resolves the sign problem, `classical' challenges in the simulation of Stoq circuits could still persist. These include the presence of an exponential number of local minima, which are inherent to NP-hard problems such as spin glasses. Such difficulties affect both AQC and QMC simulations. These obstacles are not expected to be circumvented by a universal quantum computer, that is, it is expected that $\text{NP} \not\subset \text{BQP}$, as expressed in computational complexity terms.
%---------------------------------------------------------------------------------------
%---------------------------------------------------------------------------------------
%
%        
%---------------------------------------------------------------------------------------
%---------------------------------------------------------------------------------------
 % \section{Efficiently simulating superconducting flux qubits}
 % \label{sec:FD}
\section{simulating superconducting flux circuits}

In what follows, we provide a methodology for a sign-problem free simulation of general superconducting flux circuits whose Hamiltonians are of the form,
%****************************************************************
	\begin{equation}
	\label{sec:theory-eq:Generic_Hamiltonian}
	\hat{H}(\hat{q}_1,..,\hat{q}_n,\hat{\phi}_1,...,\hat{\phi}_n) = 
	\sum_{k=1}^n ~\mu_k \hat{q}_k^2
	+\hat{V}(\hat{\phi}_1,...,\hat{\phi}_n),
	\end{equation}
	%****************************************************************
where $\hat{q}_j$ and $\hat{\phi}_k$ satisfy $\big[ \hat{q}_j,\hat{\phi}_k \big]=-i\hbar\delta_{jk}$. 
While most modern qubit devices are not necessarily given in this form~\cite{vool2017introduction}, we provide a prescription involving simple, scalable coordinate transformation, that makes them so.   
Once cast in this form, we show that they can be efficiently simulated via sign-problem-free QMC~\cite{landau:05,newman:99}, where by ``efficient'' we mean the computation will result in the thermally averaged observables that require sampling the distribution a poly($n$) number of times.
~\footnote{Of course, not much can be said about the scaling of the QMC equilibration time; even the most carefully chosen basis may still yield painfully long computation times due to the system being difficult to equilibrate, e.g., if it has a highly degenerate ground state. This is true regardless of system size, and is not unique to QMC---i.e., quantum annealers may similarly take exponential times to equilibrate.}.   

We start by considering the flux circuit Hamiltonian for $n$ rf-SQUIDs 
%****************************************************************  
	\begin{align}
    \hat{H} = & \frac{1}{2}
    \vec{Q}^{T}C^{-1}\vec{Q}+ \frac{1}{2}\left( \vec{\Phi} - \vec{\phi}_{z} \right)^{T}
    L^{-1}
    \left( \vec{\Phi} - \vec{\phi}_{z}
    \right) 
    \nonumber \\
    & - \frac{\phi_0}{2\pi} \sum_{i}^{N} I_{i} 
    \cos{\left(\frac{\pi \phi_{i}^{x}}{\phi_0}\right)}
    \cos{\left(\frac{2\pi}{\phi_0}\hat{\Phi}_i\right)},
    \label{sec:theory-eq:N-flux}
    \end{align}
%****************************************************************
with $\vec{Q}$ and $\vec{\Phi}$ being the vectors of charge and flux operators, $\hat{Q}_i$ and $\hat{\Phi}_i$, respectively. The matrices $C$ and $L$ are the capacitance and inductance matrices which are given by the circuits parameters and geometry. $I_i$ are the persistent currents, and the coaxial and transverse fluxes $\phi_z$ and $\phi_x$ serve as circuit parameters that might change during the adiabatic computation. 

We could in principle continue by discretizing the flux variables, $\hat{\phi}$, onto an equally spaced grid of unit $\Delta$ (in dimensionless flux units). This would yield a diagonal potential energy matrix of the form,
%****************************************************************
	\begin{equation}
    \sum_{IJ}\langle I |\hat{V}|J \rangle |I\rangle\langle J| = \sum_{I}^N V(i_1\Delta,...,i_n \Delta) |I\rangle\langle I|,
    \end{equation}
    %****************************************************************
where $|I\rangle = \prod_{k=1}^n |i_k\rangle=|i_1i_2...i_n\rangle$ is a direct product of the one-dimensional Dirac-delta like functions~\cite{colbert1992novel,littlejohn2002general}.  
% This yields a diagonal potential energy matrix, regardless of how complicated the functional form of $V(\phi_1,...\phi_n)$.
 
To address the kinetic energy operator we note that $\langle I|\hat{Q}_i|J \rangle \rightarrow -i\hbar\frac{d}{d\Phi_i}$ when $\Delta \rightarrow 0$. This allows us to express the derivative for a finite $\Delta$ using first order finite difference (FD), 
%****************************************************************
	\begin{equation}
	\label{sec:theory-eq:q}
    \sum_{IJ}\langle I|\hat{q}|J \rangle |I\rangle\langle J| =
    \sum_{I}^N -\frac{i\hbar}{2\Delta}
    \Big( |I+1\rangle \langle I|- |I-1\rangle \langle I|\Big). 
    \end{equation}
    %****************************************************************	
While this may not seem the natural choice at first, we note that using a more traditional basis, e.g., direct products of orthogonal polynomials, will result in positive off-diagonal elements. This is caused, in part, by the non-locality of orthogonal polynomials~\cite{halverson2015large}. Furthermore, utilizing a higher-order derivative expansion is also problematic, since it will also result in NStoq terms in the kinetic energy matrix~\cite{colbert1992novel}.
   
Obviously, Eq.~(\ref{sec:theory-eq:q}) does not immediately solve our problem, since it does not guarantee only non-positive off-diagonal matrix elements, as superconducting circuits may involve couplings between charge operators of different qubits---i.e., cross terms~\cite{harris2010experimental}, which reflect in non-diagonal matrix $C$ in Eq.~(\ref{sec:theory-eq:N-flux}). In fact, in Appendix~\ref{sec:average-sign-th} we demonstrate that if left untreated, the charge coupling terms cause a severe sign problem, i.e., the average sign will vanish exponentially fast with decreasing temperature or increasing coupling strength. 

We thus return to the continuous version of the Hamiltonian.
In order to bring the Hamiltonian in Eq.~(\ref{sec:theory-eq:N-flux}) into the general form of Eq.~(\ref{sec:theory-eq:Generic_Hamiltonian}) we transform the coordinate system in an effort to ensure that the kinetic energy only contains a sum of one-body quadratic terms. We do so by noticing that since the matrix $C$ is symmetric positive semi-definite, one can always find an orthogonal matrix $O$, such that $OCO^T=\tilde{C}$ with $\tilde{C}$ being a diagonal matrix with non negative diagonal entries. One can define $\vec{q} = O\vec{Q}$ and $\vec{\phi} = O^T\vec{\Phi}$ which serve as conjugate pair of normal-mode coordinate, owing to the fact that $O$ is orthogonal. The Hamiltonian in the new coordinates is 
%****************************************************************
    \begin{align}
     \hat{H}_{\text{flux}} = & \frac{1}{2}
     \vec{q}^{T}\tilde{C}^{-1}\vec{q}+ \frac{1}{2}\left( \vec{\phi} - \vec{\phi}_{z} \right)^{T}
     \tilde{L}^{-1}
     \left(\vec{\phi} - \vec{\phi}_{z}\right) 
      \\ & \nonumber
     - \frac{\phi_0}{2\pi} \sum_{i}^{N} I_{i} 
     \cos{\left(\frac{\pi\phi_{i}^{x}}{\phi_0}\right)}
     \cos{\left(\frac{2\pi}{\phi_0} \sum_{j}^{N} O_{ij} \hat{\phi}_j\right)}, 
     \end{align}
%****************************************************************
% To resolve this issue, we transform the coordinate system in an effort to ensure that the kinetic energy only contains a sum of one-body quadratic terms. By rotating the coordinate basis, effectively diagonalizing the capacitance matrix under the restriction that the phase space volume is preserved, we are left with the desired form. This is akin to a normal mode transformation. 
We emphasize that this normal-mode transformation is not a scaling bottleneck, since it hinges on diagonalizing (in the worst case) a matrix of dimension $n\times n$, with $n$ being the number of charge operators. This procedure allows for a generic way to simulate the bulk of circuits used throughout AQC~\cite{krantz2019quantum} by mapping the continuous circuit Hamiltonian for the device to the discretized Hamiltonian given by,
%****************************************************************
	\begin{align}
    \hat{H}_{\textsc{fd}} = &
    \sum_I \sum_{k=1}^n -\frac{\mu_k\hbar^2}{\Delta^2}
    \Big( |I_k^{(+)}\rangle \langle I|-2|I\rangle \langle I|     + |I_k^{(-)}\rangle \langle I|\Big)
        \nonumber \\
    &+ \sum_{I} V(i_1\Delta,...,i_n \Delta) |I\rangle\langle I|, 
    \label{sec:theory-eq:discrete_H}
    \end{align}    %**************************************************************** 
where $\mu_k$ are half the diagonal elements of $\tilde{C}^{-1}$, and $|I_k^{(\pm)}\rangle=|i_1 ... i_{k-1}\rangle \otimes |i_{k}\pm1\rangle \otimes |i_{k+1}...i_n\rangle$, obviating the qubit approximation entirely. 
% ---------------------------------------------------------------------------------------
% ---------------------------------------------------------------------------------------

% ---------------------------------------------------------------------------------------
% ---------------------------------------------------------------------------------------

\section{Simulating non-stoquastic flux qubit Hamiltonians: Two qubits}
\subsection{System}
\begin{figure}
	\centerline{\includegraphics[width=.45\textwidth]{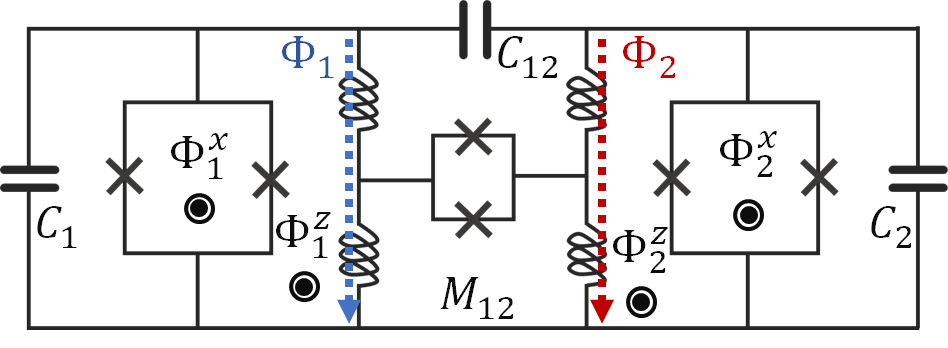}}
	\caption{Circuit schematic for the non-stoquastic two qubit rf-SQUID~\cite{dwaveNonStoq}. The two superconducting Josephson 		junctions are coupled both capacitively and inductively.}
	\label{fig:circuit}
	\end{figure}

For a macroscopic system, such as a superconducting flux circuit, to be a viable choice as a quantum information device, one must be able to isolate and address a subspace generated by two states per qubit. 
Because of this, rf-SQUIDs have become the cornerstone of modern quantum computing~\cite{fagaly2006superconducting}. They can be modeled as an anharmoic oscillator, where the two lowest energy states are spectroscopically accessible through some control process. Moreover, these two states can be mapped to an analogous two-level spin system where traditional quantum computing operations may be carried out. This is all while being easily fabricated through well understood processes~\cite{fagaly2006superconducting}. One such rf-SQUID device, the schematic for which is shown in Fig.~\ref{fig:circuit}, and whose Hamiltonian is given by Eq.~(\ref{sec:theory-eq:N-flux}) for $n=2$, 
	% %****************************************************************  
 %   \begin{equation}
 %   \hat{H}_c=
 %     \frac{\hat{Q}^2_1}{2\tilde{C}_1}
 %   + \frac{\hat{Q}^2_2}{2\tilde{C}_2}
 %   + \frac{C_{12}~\hat{Q}_1 \hat{Q}_2}{C_1 C_2+C_{12}(C_1+C_2)} \nonumber
 %   \end{equation}
 %   \begin{equation}
 %   + \frac{(\hat{\Phi}_1-\phi_1^z)^2}{2L_1}
 %   + \frac{(\hat{\Phi}_2-\phi_2^z)^2}{2L_2}
 %   + \frac{M_{12}(\hat{\Phi}_1-\phi_1^z)(\hat{\Phi}_2-\phi_2^z)}{L_1 L_2} \nonumber
 %   \end{equation}
 %   \begin{equation}
 %   \label{sec:ham-eq:Hamiltonian}
 %   - \frac{\phi_0I_1}{2\pi} \cos\Big[\frac{\pi \phi_1^x}{\phi_0}\Big]
 %   \cos\Big[\frac{2\pi}{\phi_0}\hat{\Phi}_1  \Big]
 %   - \frac{\phi_0I_1}{2\pi} \cos\Big[\frac{\pi \phi_2^x}{\phi_0}\Big]
 %   \cos\Big[\frac{2\pi}{\phi_0}\hat{\Phi}_2  \Big].
 %   \end{equation}
 %   	%****************************************************************
has been specifically designed to be NStoq in its qubit representation, owing to the novel couplings between the the two qubits (note that $\phi_0=\pi \hbar/e$ with $\hbar = e = 1$ throughout)~\cite{dwaveNonStoq}. 
	%~~~~~~~~~~~~~~~~~~~~~~~~~~~~~~~~~~~~~~
	%~~~~~~~~~~~~~~~~~~~~~~~~~~~~~~~~~~~~~~
We present here a detailed study of how the method outlined in the previous section can be used to simulate this circuit efficiently. 

\begin{figure}
	\centerline{\includegraphics[width=.45\textwidth]{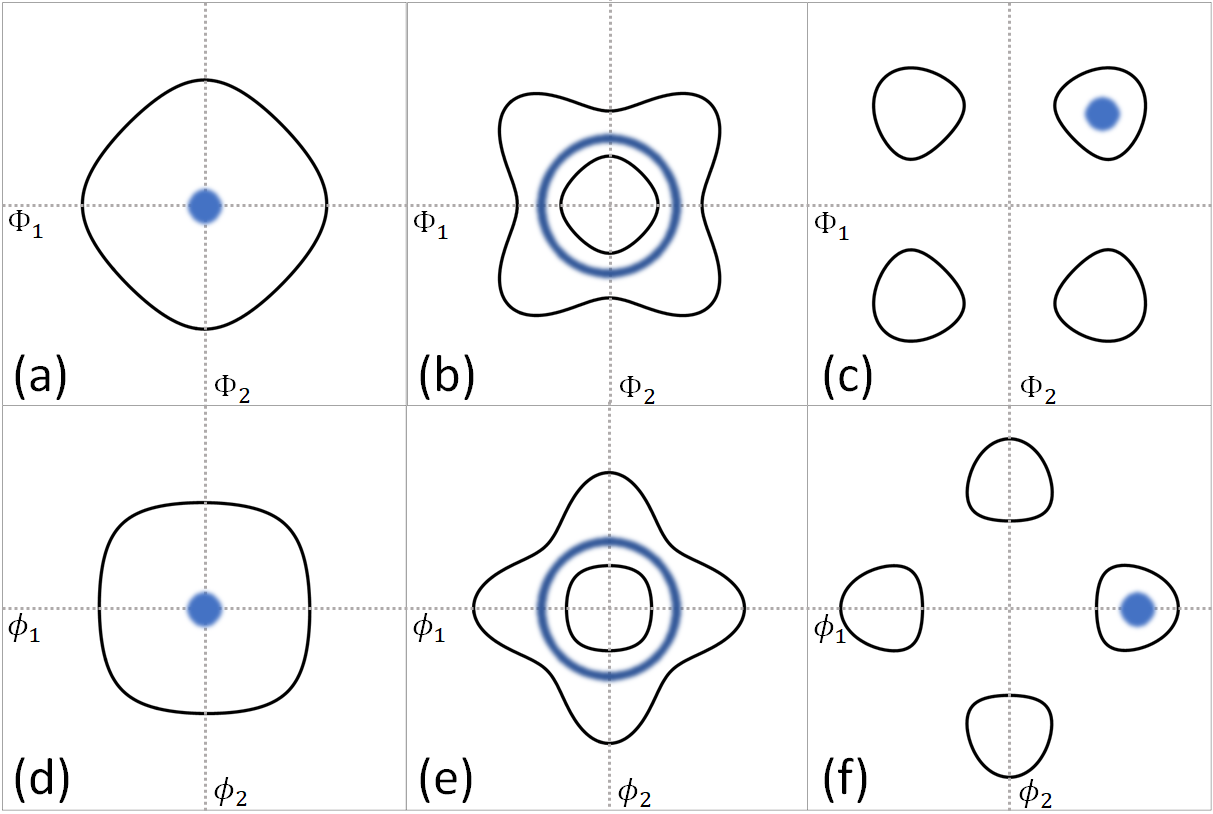}}
	\caption{Schematic overview of the change in the potential energy of the two qubit circuit Hamiltonian during the annealing 		process. (a) Single contour of the potential energy at $\phi_x=0$. The dot represents the majority of the probability density 		of the ground state wavefunction. (b) Single contour of the potential energy at $\phi_x=3\pi/4$. The solid ring represents the 		majority of the probability density of the ground state wavefunction, which is delocalized due to the presence of a phase 			transition. (c) Single contour of the potential energy at $\phi_x=\pi$. The dot represents the majority of the probability 			density of the ground state wavefunction, which is in the well associated with the $|00\rangle$ qubit state. (d)-(f) Same as 		(a), (b), and (c) but after the normal mode transformation.}
	\label{fig:anneal}
	\end{figure}

The values of the fixed parameters in the above Hamiltonian are given in Table~\ref{table:params}. 
	%~~~~~~~~~~~~~~~~~~~~~~~~~~~~~~~~~~~~~~
	\begin{table}
	\begin{tabular}{ccccc}
	\hline
	Parameter     & &  Qubit 1 & & Qubit 2\\
	\hline
	\hline    
	$L$ (pH)      & &  231.9   & &  239.1  \\
	\hline
	$C$ (fF)      & &  119.5   & &  116.4  \\
	\hline
	$I$ ($\mu$A)  & &  3.227   & &  3.157  \\
	\hline
	$C_{12}$ (fF) & &  \multicolumn{3}{c}{132.0}  \\
	\hline
	\end{tabular}
	\caption{Fixed circuit parameters for the coupled rf-SQUID system. See Ref.~\cite{dwaveNonStoq} for more details.}
	\label{table:params}
	\end{table}
	%~~~~~~~~~~~~~~~~~~~~~~~~~~~~~~~~~~~~~~
The parameters not listed in Table~\ref{table:params} are adjustable device parameters used for initializing and performing the annealing process. To initialize the system, the coaxial fluxes, $\phi_1^{z}$ and $\phi_2^z$, are set to the values associated with the problem Hamiltonian. These flux values define the final state to be measured at the end of the annealing. The transverse fluxes, $\phi_1^x=\phi_2^x=\phi_x$, are then varied slowly, transforming the system from a single anharmonic well at $\phi_x = 0$ to a set of four wells at $\phi_x = \pi$~\cite{DWave-entanglement}. These four wells define the four possible 2-qubit states, the lowest of which being the ground state of the final system and the solution to the problem Hamiltonian. This process is shown schematically in Fig.~\ref{fig:anneal}.
	%~~~~~~~~~~~~~~~~~~~~~~~~~~~~~~~~~~~~~~
	%~~~~~~~~~~~~~~~~~~~~~~~~~~~~~~~~~~~~~~

As was outlined in the previous section, we cannot work directly with Eq.~(\ref{sec:theory-eq:N-flux}), owing to the first order charge coupling ($\hat{Q}_1 \hat{Q}_2$). Instead we transform to the normal mode coordinates, which for the two rf-SQUIDs system can be done analytically and given by 
	%****************************************************************
	\begin{equation}
	\hat{q}_{1(2)} = \Bigg(\frac{1}{8\sqrt{\omega_1 \omega_2}}\Bigg)^{1/2}
	\Big(
	\sqrt{\omega_1}~\hat{u}_1 \mp  \sqrt{\omega_2}~\hat{u}_2
	\Big)	
    \label{eq:two_qubits_transformation}
	\end{equation}
 
	%****************************************************************
and,
	%****************************************************************
	\begin{equation}
	\hat{\phi}_{1(2)} = \Bigg(\frac{2}{\sqrt{\omega_1 \omega_2}}\Bigg)^{1/2}
	\Big(
	\sqrt{\omega_2}~\hat{v}_1 \mp  \sqrt{\omega_1}~\hat{v}_2
	\Big),	
	\end{equation}
	%****************************************************************
The dimensionless operators, $\hat{u}$ and $\hat{v}$, are defined as \hbox{$\hat{Q}_i = \sqrt{\frac{\hbar \omega_i}{C^{-1}_{ii}}}~\hat{u}_i$} and \hbox{$\hat{\Phi}_i = \sqrt{ \frac{\hbar C^{-1}_{ii}}{\omega_i}} ~\hat{v}_i + \phi_i^z$}, with \hbox{$\omega_i^2= C^{-1}_{ii} / L_{ii}$}. This yields kinetic and potential operators of the form,
%****************************************************************
   \begin{equation}
   \hat{T}(\hat{q}_1,\hat{q}_2) =C_{+}~\hat{q}^2_1+C_{-}~\hat{q}^2_2,
   \end{equation}
   %****************************************************************
 and,   %****************************************************************
   \begin{align}
   \hat{V}(\hat{\phi}_1,\hat{\phi}_2) = & L_{-}~\hat{\phi}^2_1+L_{+}~\hat{\phi}^2_2 
   + L_{12}~\hat{\phi}_1\hat{\phi}_2
   \\ \nonumber &
   + E_1(\phi_x)\cos\Big[ \frac{2\pi}{\phi_0}\big( \Omega_1(\hat{\phi}_1+\hat{\phi}_2)+\phi_1^z\big)\Big]
   \\ \nonumber &
   + E_2(\phi_x)\cos\Big[ \frac{2\pi}{\phi_0}\big( \Omega_2(\hat{\phi}_1-\hat{\phi}_2)+\phi_2^z\big)\Big],
   \end{align}
%****************************************************************
the coefficients of which are presented in Table~\ref{table:comps}. 
With this the Hamiltonian acquires the stoquastic form of Eq.~(\ref{sec:theory-eq:discrete_H}).

\begin{table}[!htb]
\begin{tabular}{ccc}
\hline
Coefficient     & &  Expression\\
\hline
\hline 
$\tilde{C}_{1(2)}$ & &  $C_{1(2)}+\frac{C_{12}C_{2(1)}}{C_{2(1)}+C_{12}}$ \\
	\hline
$C_{\pm}$          & &  $2\hbar\sqrt{\omega_1 \omega_2}\Big(1\mp\frac{C_{12} \sqrt{ \tilde{C}_1 \tilde{C}_2}}{C_1 C_2+C_{12}(C_1+C_2)} \Big)$ \\
     \hline   
$L_{\pm}$          & &  $  \frac{\hbar}{16\sqrt{\omega_1 \omega_2}}\Big(
\omega_1^2+\omega_2^2 \mp \frac{2 M_{12}}{L_l L_2\tilde{C}_1\tilde{C}_2} \Big)$  \\
      \hline
$L_{12}$           & &  $\frac{\hbar}{4\sqrt{\omega_1 \omega_2}}\big(\omega_1^2-\omega_2^2 \big) $ \\
	\hline
$E_i(\phi_x)$      & & $\frac{\phi_0}{2\pi}I_i \cos\big[ \frac{\pi}{\phi_0}\phi_x  \big]$  \\
	\hline
$\Omega_{i}$      & &  $\Big(\frac{\hbar}{8\tilde{C}_{i}\sqrt{\omega_1 \omega_2}}\Big)^{1/2}$   \\
      \hline
\end{tabular}
\caption{Coefficients for the two qubit rf-SQUID circuit normal mode Hamiltonian.}
\label{table:comps}
\end{table}

%---------------------------------------------------------------------------------------
%---------------------------------------------------------------------------------------
%
%
%---------------------------------------------------------------------------------------
%---------------------------------------------------------------------------------------

\subsection{Results} 
%\subsection{Two Qubits system}
We simulate the circuit outlined in the previous subsection, Eq.~(\ref{sec:theory-eq:discrete_H}), by employing off-diagonal expansion QMC (ODE QMC)~\cite{ODE,ODE2,pmr} (The reader is referred to Appendix~\ref{sec:exp_details} for technical details). Using this, we can compute the thermal average of observable quantities by $\langle \hat{A} \rangle = \textsf{Tr}[\hat{A} e^{-\beta \hat{H}}]/\textsf{Tr}[e^{-\beta \hat{H}}]$, where $\beta=1/kT$ (here we chose $T=12$~mK to reflect actual device temperatures). The quantity of interest for superconducting flux qubits and AQC is the persistent current, $\hat{I}_{1(2)}^z$, the expectation value of which can be expressed as,
%****************************************************************
  	\begin{align}
  	\label{sec:results-eq:I}
  	\langle \hat{I}_{1(2)}^z \rangle = & -\left\langle\frac{\partial \hat{H}}{\partial \phi_{1(2)}^z}\right\rangle =
    \\ \nonumber
  	& \frac{1}{L_{1(2)}}\left(\langle\hat{\Phi}_{1(2)}\rangle-\phi_{1(2)}^z\right)
  	+\frac{M_{12}}{L_1 L_1}\left(\langle\hat{\Phi}_{2(1)}\rangle-\phi_{2(1)}^z\right).
  	\end{align}  	%****************************************************************
The above constitutes the ``qubit'' in superconducting flux qubit device, since the quantized current (or more appropriately, the direction of the quantized current) is what maps to the spin up (or down) in the qubit model, analogous to the eigenstates of the Pauli-z operator~\cite{Kjaergaard2020}. In other words, this is the quantity that is measured by the hardware, and a clockwise (or counterclockwise) current yields a positive (or negative) current measurement, which is interpreted as a qubit state of $|0\rangle$ (or $|1\rangle$). 

\begin{figure}[!htb]
	\centerline{\includegraphics[width=.45\textwidth]{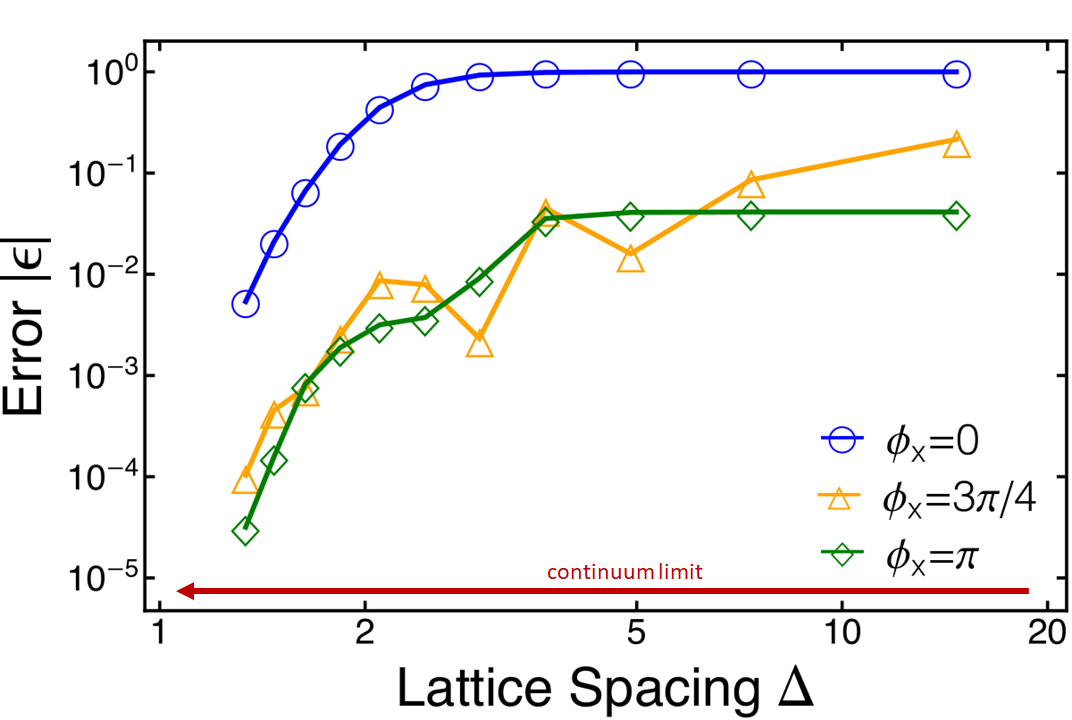}}
	\caption{Magnitude of the relative error of the persistent current as a function of the convergence parameter 
	$\Delta$. The current was computed using   $\phi_1^z=0.1~m\phi_0$ and  $\phi_2^z=0.9~m \phi_0$ at three different points in 	the anneal. 
	}
	\label{fig:Delta}
	\end{figure}

Our algorithm has only one extraneous parameter, the grid spacing $\Delta$. It is important to make sure that in the limit $\Delta\rightarrow 0$, the discretized model approaches the continuous Hamiltonian. 
This is presented in Fig.~\ref{fig:Delta} which shows the relative error for $\langle \hat{I}_1^z\rangle$ as a function of $\Delta$ computed using exact diagonalization, confirming our expectations that our simulations recover the results of the continuous model for small enough $\Delta$. 
	%~~~~~~~~~~~~~~~~~~~~~~~~~~~~~~~~~~~~~~
	%~~~~~~~~~~~~~~~~~~~~~~~~~~~~~~~~~~~~~~
	
\begin{figure}[!htb]
	\centerline{\includegraphics[width=.48\textwidth]{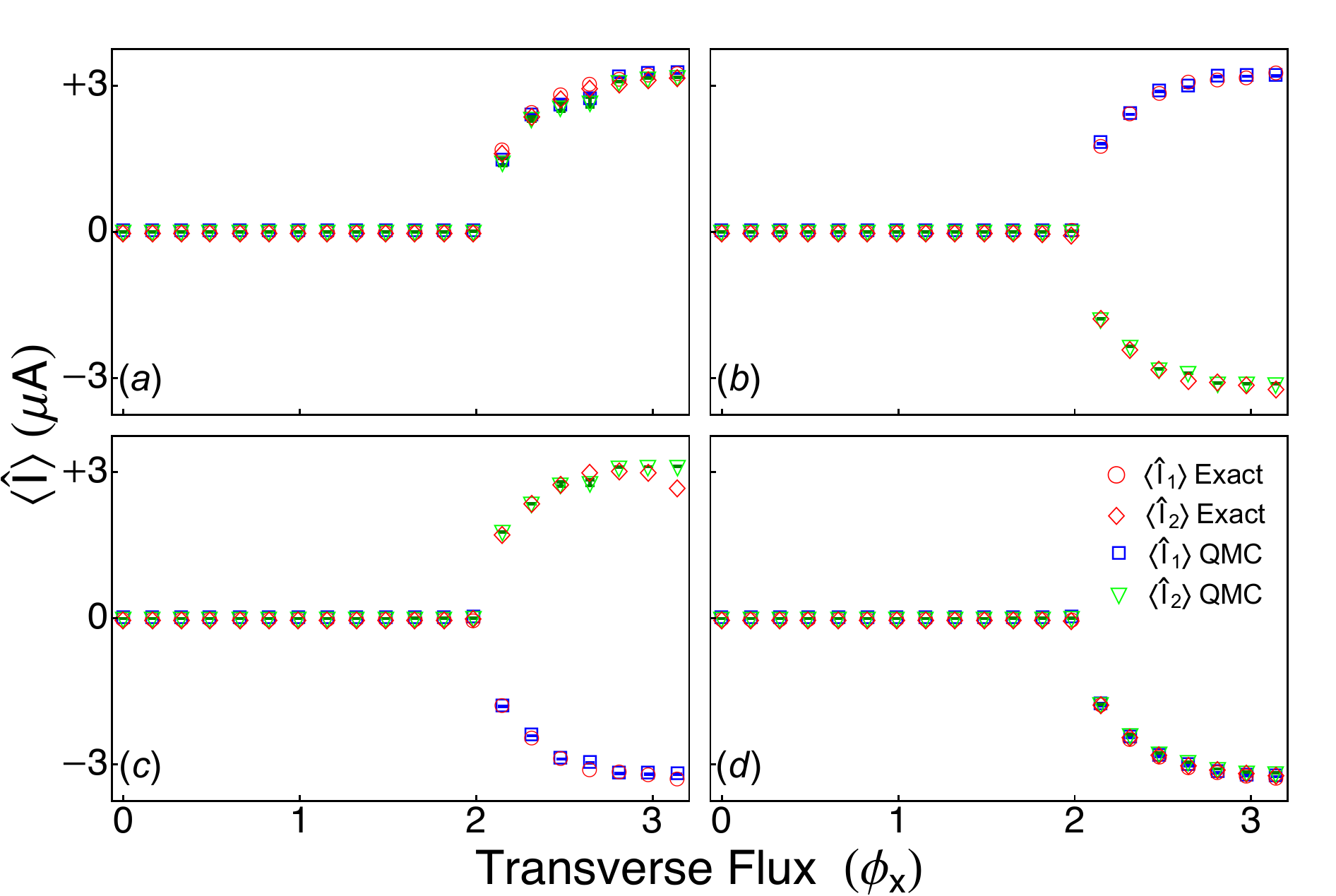}}
	\caption{Persistent current calculated for both exact diagonalization and ODE QMC at various $\phi_x$ points throughout the 		annealing process. (a) Anneal for the $|00\rangle$ qubit output. $\phi_1^z=0.1~\text{m}\phi_0$ and  $\phi_2^z=0.9~\text{m} \phi_0$ were used 	as the problem Hamiltonian values. (b) $|01\rangle$, $\phi_1^z=0.5$ and  $\phi_2^z=-0.5$ (c) $|10\rangle$, $\phi_1^z=-0.4$ and  $\phi_2^z=0.25$ (d) $|11\rangle$, $\phi_1^z=-0.25 $ and  $\phi_2^z=-0.25$.}
	\label{fig:IZ}
	\end{figure}

To demonstrate that we can, in fact, simulate the circuit, we computed four separate anneals for four sample problem Hamiltonians, each corresponding to the four possible 2-qubit readouts---$~|00\rangle, |01\rangle,|10\rangle$, and $|11\rangle$. We dictate which of the final states we desire by setting the coaxial fluxes, $\phi_1^z$ and $\phi_2^z$, to either positive or negative values depending on the desired state. The magnitude of these chosen values would come from the problem Hamiltonian one would be attempting to solve using AQC. For the purposes here, they were chosen arbitrarily to showcase the versatility of our technique. This is illustrated in Fig.~\ref{fig:IZ}. 

We can relate the change in $\hat{I}_z$ shown in Fig.~\ref{fig:IZ} back to the schematic shown in Fig.~\ref{fig:anneal} by noting again that each of the four qubits states correspond to a well in one of the four Cartesian quadrants of the $(\hat{\Phi}_1,\hat{\Phi}_2)$ plane. These qubit states also correspond to a set of signs of $I_1^z$ and $I_2^z$. For example, setting both $\phi_1^z$ and $\phi_2^z$ to positive values results in the well in the $(+,+)$ quadrant being the lowest in energy, as well as  in $I_1^z$ and $I_2^z$ both having a positive sign, as in Fig.~\ref{fig:IZ}(a). Expressed differently, Fig.~\ref{fig:IZ} illustrates that the sign of the current readout corresponds to the quadrant in which the lowest energy well is located.   
	%~~~~~~~~~~~~~~~~~~~~~~~~~~~~~~~~~~~~~~

It is interesting to note that the qubit representations of the circuits simulated above are virtually unsimulable by QMC due to the severe sign problem caused by the NStoq terms that arise from the reduction of the circuit~\cite{gupta2019,dwaveNonStoq}.

\subsection{Larger System sizes}

\begin{figure}[H]
    \centerline{\includegraphics[width=.5\textwidth]{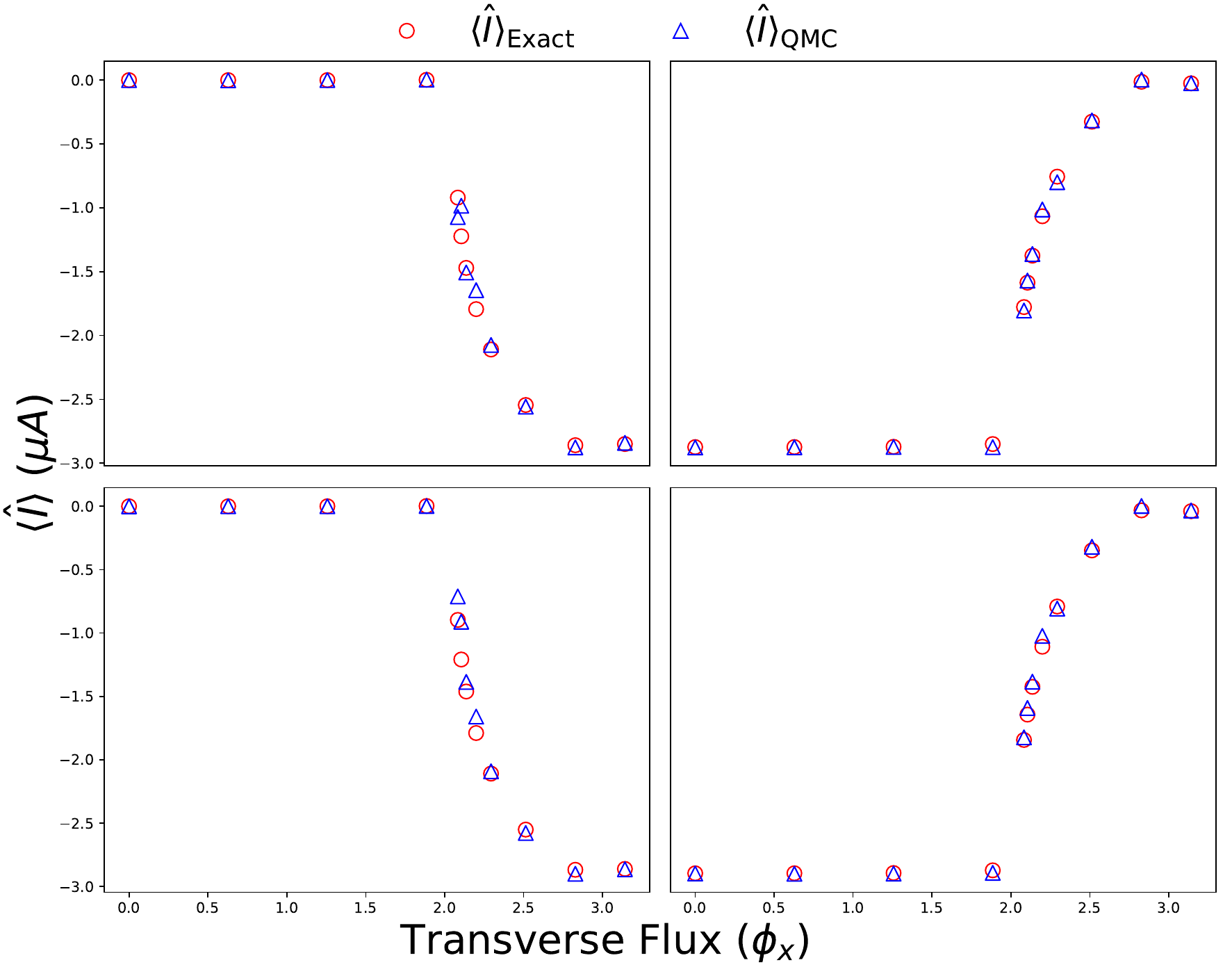}}
	\caption{Persistent current calculated for both exact diagonalization and QMC as a function of the transverse flux $\phi_x$ for four rf-SQUIDs system arranged on a square lattice. 
	}
	\label{fig:four_qubits}
	\end{figure}

%~~~~~~~~~~~~~~~~~~~~~~~~~~~~~~~~~~~~~~
Our approach allows to address much larger systems.
As an example, in Fig.~\ref{fig:four_qubits} we present the results of a simulation of a system of four rf-SQUIDs arranged on a square lattice (with nearest-neighbor cross-couplings) with parameters values listed in Table.~\ref{table:four_params}. 
We also simulated linear chains of six and eight rf-SQUID (with nearest-neighbor cross-couplings).
Due to the larger system size, we present the results for a given set of parameters and coaxial and transverse flux in Table.~\ref{table:six_params} and Table.~\ref{table:eight_params}, respectively. These include the average relative error, which is defined as $\frac{1}{n} \sum_{i}^{n} \frac{\left< I_\text{Exact}\right>_{i} - \left<I_\text{QMC}\right>_{i}}{\left< I_\text{Exact}\right>_{i}}$ with $n$ being the number of rf-SQUIDs.

	%~~~~~~~~~~~~~~~~~~~~~~~~~~~~~~~~~~~~~~
	\begin{table}[!h]
	\begin{tabular}{ccccc}
	\hline
	Parameter     &  Qubit 1 & Qubit 2 & Qubit 3  & Qubit 4 \\
	\hline
	\hline    
	$L$ (pH)      &  231.9   &  232  &  231   &  230 \\
	\hline
	$C$ (fF)      &  119.5   &  120  &  120.5   &  121 \\
	\hline
	$I$ ($\mu$A)  &  3.227   &  3.227  &  3.227   &  3.227 \\
	\hline
    $\phi^{z}$ (m$\phi_0$)  &  0.1   &  0.1  &  0.1   &  0.9 \\
	\hline
	$C_{ij}$ (fF) &  \multicolumn{2}{c}{132.0}  \\
	\hline
     $L_{ij}$ (pH) &  \multicolumn{2}{c}{0.2}  \\
    \hline
	\end{tabular}
	\caption{Circuit parameters for the system of four rf-SQUIDs arranged in a square. Simulation results are plotted in Fig.~\ref{fig:four_qubits}}.
	\label{table:four_params}
	\end{table}
	%~~~~~~~~~~~~~~~~~~~~~~~~~~~~~~~~~~~~~~

	%~~~~~~~~~~~~~~~~~~~~~~~~~~~~~~~~~~~~~~

	\begin{table}[!h]
	\begin{tabular}{ccccc}
	\hline
	Parameter     &  Qubit 1 & Qubit 2 & Qubit 3  \\ 
	\hline
	\hline    
	$L$ (pH)      &  231.9   &  232  &  231  \\
	\hline
	$C$ (fF)      &  119.5   &  120  &  120.5  \\
	\hline
	$I$ ($\mu$A)  &  3.227   &  3.227  &  3.227 \\
	\hline
    $\phi^{z}$ (m$\phi_0$)  &  0.9   &  0.9  &  0.9  \\
     \hline
     \hline

	Parameter     &  Qubit 4 & Qubit 5 & Qubit 6  \\ 
	\hline
	\hline    
	$L$ (pH)      &  230   &  230.5  &  231.3  \\
	\hline
	$C$ (fF)      &  121   &  120.7  &  119.8  \\
	\hline
	$I$ ($\mu$A)  &  3.227   &  3.227  &  3.227 \\
	\hline
    $\phi^{z}$ (m$\phi_0$)  &  0.9   &  0.9  &  0.9  \\
    \hline
    $C_{ij}$ (fF) &  \multicolumn{2}{c}{132.0}  \\
	\hline
     $L_{ij}$ (pH) &  \multicolumn{2}{c}{0.2}  \\
    \hline
    \hline

	Results     & &  Qubit 1 & Qubit 2 & Qubit 3  \\ 
    \hline

    $\left< \hat{I}_{\text{Exact}} (\mu A) \right>$ &  0.960 & 0.962  &  0.937 \\
    \hline
    $\left< \hat{I}_{\text{QMC}} (\mu A) \right>$  &  0.965   &  0.984  &  0.943   \\
    \hline
	\hline  
 
    Results     &  Qubit 4 & Qubit 5 & Qubit 6  \\ 
    \hline

    $\left< \hat{I}_{\text{Exact}} (\mu A) \right>$ &  0.9105 & 0.924  &  0.945 \\
    \hline
    $\left< \hat{I}_{\text{QMC}} (\mu A) \right>$  &  0.882   &  0.835  &  0.947   \\
    \hline
    \hline
    Relative Error &   \multicolumn{2}{c}{0.027}  \\

	\end{tabular}
	\caption{Circuit parameters and simulation results for a linear chain of six rf-SQUIDs.}
	\label{table:six_params}
	\end{table}
 
   	%~~~~~~~~~~~~~~~~~~~~~~~~~~~~~~~~~~~~~~
    
	%~~~~~~~~~~~~~~~~~~~~~~~~~~~~~~~~~~~~~~

	\begin{table}[!htb]
	\begin{tabular}{ccccc}
	\hline
	Parameter     &  Qubit 1 & Qubit 2 & Qubit 3  & Qubit 4\\ 
	\hline
	\hline    
	$L$ (pH)      &  231.9   &  232  &  231  &  230 \\
	\hline
	$C$ (fF)      &  119.5   &  120  &  120.5  &  121 \\
	\hline
	$I$ ($\mu$A)  &  3.227   &  3.227  &  3.227 &  3.227\\
	\hline
    $\phi^{z}$ (m$\phi_0$)  &  0.9   &  0.9  &  0.9  &  0.9 \\
     \hline
     \hline
    
	Parameter     &  Qubit 5 & Qubit 6 & Qubit 7  & Qubit 8\\ 
	\hline
	\hline    
	$L$ (pH)      &  230.5   &  231.3  &  230.6  &  231.1 \\
	\hline
	$C$ (fF)      &  120.7   &  119.8  &  119.9  &  120.8 \\
	\hline
	$I$ ($\mu$A)  &  3.227   &  3.227  &  3.227 &  3.227\\
	\hline
    $\phi^{z}$ (m$\phi_0$)  &  0.9   &  0.9  &  0.9  &  0.9\\
    \hline
    $C_{ij}$ (fF) &  \multicolumn{2}{c}{132.0}  \\
	\hline
     $L_{ij}$ (pH) &  \multicolumn{2}{c}{0.2}  \\
    \hline
    \hline

	Results     &  Qubit 1 & Qubit 2 & Qubit 3  & Qubit 4\\ 
    \hline

    $\left< \hat{I}_{\text{Exact}} (\mu A) \right>$ &  0.971 & 0.973  &  0.949 &  0.923\\
    \hline
    $\left< \hat{I}_{\text{QMC}} (\mu A) \right>$  &  0.987   &  1.055  &  0.978 &  0.971 \\
    \hline
	\hline  
 
    Results     &  Qubit 5 & Qubit 6 & Qubit 7  & Qubit 8 \\ 
    \hline

    $\left< \hat{I}_{\text{Exact}} (\mu A) \right>$ &  0.936 & 0.956  &  0.939 &  0.951\\
    \hline
    $\left< \hat{I}_{\text{QMC}} (\mu A) \right>$  &  0.920   &  0.866  &  0.991   &  0.940\\
    \hline
    \hline
    Relative Error &  \multicolumn{2}{c}{0.045}  \\

	\end{tabular}
	\caption{Circuit parameters and simulation results for a linear chain of eight rf-SQUIDs.}
	\label{table:eight_params}
	\end{table}
 
   	%~~~~~~~~~~~~~~~~~~~~~~~~~~~~~~~~~~~~~~

%---------------------------------------------------------------------------------------
%---------------------------------------------------------------------------------------

\section{Conclusions}

In this study we devised a prescription to make seemingly NStoq superconducting flux circuits amendable to sign-problem free QMC techniques. 
The importance of our work lies mainly in its implications for approximating superconducting circuits as qubit model, and what information about the circuit can be gleaned from simulations of the model. Our results stand in stark contrast with recent and ongoing efforts to construct circuits of this type~\cite{dwaveNonStoq}. If indeed superconducting flux circuits---even those circuits whose qubit representations are NStoq---do not suffer from the QMC sign-problem, then the implications are that flux circuits are not likely to generate genuine non-stoquastic dynamics, which is an important ingredient for both AQC universality and quantum supremacy. 

While the focus of our work has been superconducting flux qubit devices, we note that our analysis equally applies for any ``traditional'' qubit circuit. We note that while we were able to avoid the sign-problem, there are still classical obstacles in the thermalization of a general flux circuit that prevent scalability due to the glassy nature of the problem. Regardless, our conclusions remain the same, since this type of hardness has nothing to do with quantum supremacy, and those obstacles are expected to be present in the AQC itself, thus showing no overhead with respect to the physical experiment. 

% Our conclusions are predicated on the grid spacing being constant as a function of system size, which is expected since the grid spacing is proportional to the scale of the distance between wells in the flux hyperspace, which are relatively constant with respect to system size. 

%We conclude that for AQC to live up to its full potential, one must look for entirely new types of flux qubit systems.
The method laid out here is predicated on the limitation that the charge part of the Hamiltonian can be made Cartesian via some sort of transformation. Therefore, an important question that arises from our work is: What modifications to the Hamiltonians of superconducting flux circuits can be made for these systems to truly exhibit non-stoquasticity at the circuit level? One obvious choice is to introduce an external vector potential; however, this will only results in an offset charge operator, which can, most likely, be gauged away via a coordinate transformation. Currently, the most likely candidate to produce a truly NStoq circuit is a Hamiltonian that couples the charge and flux operators. This is precisely what is being explored by the bifluxon qubit system~\cite{PRXQuantum.1.010307}. We conclude that AQC would benefit greatly from more effort being put into the study of systems of this type in order to achieve this long sought after goal---mainly, a truly unsimulable circuit Hamiltonian.     

An interesting conjecture that follows from our results is that every NStoq (and thus non-simulable) spin model  is a low-energy approximation of a higher-dimensional Stoq model. If this is the case and an efficient construction of such higher-dimensional models exists, that might be a path towards resolving the infamous QMC sign problem~\cite{marvian:2018,gupta2019}.
\begin{acknowledgments}
We would like to extend a special thanks to David Ferguson for many insightful discussions. We would also like to thank Tameem Albash, Elizabeth Crosson, Mostafa Khezri and Milad Marvian for valuable comments. 
The research is based upon work (partially) supported by the Office of
the Director of National Intelligence (ODNI), Intelligence Advanced
Research Projects Activity (IARPA) and the Defense Advanced Research Projects Agency (DARPA), via the U.S. Army Research Office
contract W911NF-17-C-0050. The views and conclusions contained herein are
those of the authors and should not be interpreted as necessarily
representing the official policies or endorsements, either expressed or
implied, of the ODNI, IARPA, DARPA, or the U.S. Government. The U.S. Government
is authorized to reproduce and distribute reprints for Governmental
purposes notwithstanding any copyright annotation thereon. H.K.N. and M.G. acknowledge support
by the Israel Science Foundation
(ISF) and the Directorate for Defense Research and Development
(DDR\&D) through Grant No. 3427/21, the
ISF grant No. 1113/23, and the US-Israel Binational Science
Foundation (BSF) through Grant No. 2020072.
%thanks to the U.S.-Israel Binational Science Foundation (Grant No. 2016224) and the Israel Science Foundation (Grant No. 227/15). 
\end{acknowledgments}
%---------------------------------------------------------------------------------------
%---------------------------------------------------------------------------------------

%---------------------------------------------------------------------------------------

\appendix

\section{Average sign for the two coupled rf-SQUIDs} \label{sec:average-sign-th}
Consider an Hamiltonian of the form of Eq.~(\ref{sec:theory-eq:N-flux}),
%\begin{equation}
%    H = QC^{-1}Q + U\left( \Phi \right),
%\end{equation}
with the flux operator discretized as in the main text to a basis states $\left| I\right>$, with $I\in\{ 1,...,N \}$  and consequentially, the charge operator is approximated by finite difference, as in Eq.~(\ref{sec:theory-eq:q}),
%\begin{equation}
%    Q \approx -\frac{i\hbar}{2\Delta} \sum_{I}^{N} \left| I \right> \left< I + 1 \right| - \left| I \right> \left< I - 1 \right|,
%\end{equation}
with $\Delta$ being the flux discretization scale. 
In order to quantify the severity of the sign problem we first decompose $H$ into a de-signed part, $H_D$ which is a stoquastic Hamiltonian, and a signed part $H_S$, which consists of all of the off diagonal positive terms
\begin{equation}
    H_D \equiv H - 2H_S,
    \nonumber
\end{equation}
with 
\begin{equation} 
H_S=\begin{cases}
 H_{ij} \ \   \text{if} \  H_{ij} > 0 \ \text{and} \  i\neq j \\
0 \ \ \ \ \ \text{otherwise.}
\end{cases}
\end{equation}
Defining $Z$ and $Z_{D}$ to be the partition functions of the Hamiltonian and the de-signed Hamiltonian respectively, the average sign can be defined \cite{Hangleiter_2020} for various QMC schemes by 
\begin{equation}
    s \equiv \frac{Z}{Z_D}.
\end{equation}
in order to be able to evaluate it we will prove the following lemma 
\newtheorem{theorem}{Theorem}[section]
\newtheorem{lemma}[theorem]{Lemma}
\begin{lemma}
Given an Hamiltonian $H$ and a decomposition of it to a de-signed part $H_D$ and a signed part $H_S$
the average signed satisfies 
\begin{equation}
    s \leq e^{-2 \beta \left\langle H_S \right\rangle},
\end{equation}
with $\beta$ being the inverse temperature and $\left\langle H_S \right\rangle \equiv \frac{1}{Z}\Tr \left(H_S e^{-\beta H} \right)$.
\end{lemma}
To prove the lemma we use Gibbs inequality \cite{Kardar_2007}, which states that, given two Hamiltonians, $H_A$ and $H_B$, then, $\ln{Z_{B}} \geq \ln{Z_{A}} + \beta \left< H_{A} - H_{B}\right>_{A}$, and substitue $H_{A} = H$ and $H_{B} = H_{D}$.

Consider for simplicity the Hamiltonian for two coupled rf-SQUIDs. Denoting the inverse capacitance matrix by $\tilde{C}_{ij} \equiv \left(C^{-1}\right)_{ij} $, the signed part of $H$ is
% \begin{equation}
%     A = \frac{\hbar^2}{4\tilde{C}_{12}\Delta^{2}} \sum_{I_1, I_2} \left| I_{1},I_{2} \right> \left< I_{1} + 1,I_{2} + 1 \right| + \left| I_{1},I_{1} \right> \left< I_{1} - 1, I_{2} - 1\right|.
% \end{equation}
\begin{align}
    H_S = \frac{\hbar^2}{4\tilde{C}_{12}\Delta^{2}} \sum_{I_1, I_2} & \Big( 
    \left| I_{1},I_{2} \right\rangle \left\langle I_{1} + 1,I_{2} + 1 \right| +  \notag \\
    & \left| I_{1},I_{2} \right\rangle \left\langle I_{1} - 1, I_{2} - 1\right| \Big).
\end{align}
The action of $H_S$ on a general state defined over the discrete basis $\left| \Psi\right\rangle = \sum_{I_{1},I_{2}} \psi(\Phi_{I_{1}},\Phi_{I_{2}})\left|I_{1},I_{2}\right\rangle$ is given by 
% \begin{equation}
% \begin{split}
%     A\left|f\right> =  \frac{\hbar^2}{4\tilde{C}_{12}\Delta^{2}} \sum_{I_1, I_2} 
%     \left(f(\Phi_{I_{1}}+\Delta,\Phi_{I_{2}}+\Delta) + \\ f(\Phi_{I_{1}}-\Delta,\Phi_{I_{2}}-\Delta)\right) \left|I_{1},I_{2} \right>, 
% \end{split}
% \end{equation}
\begin{align}
    H_S\left|\Psi\right\rangle =  \frac{\hbar^2}{4\tilde{C}_{12}\Delta^{2}} \sum_{I_1, I_2} & 
    \Big(\psi(\Phi_{I_{1}}+\Delta,\Phi_{I_{2}}+\Delta) +  \\ 
    & \psi(\Phi_{I_{1}}-\Delta,\Phi_{I_{2}}-\Delta)\Big) \left|I_{1},I_{2} \right\rangle. \notag
\end{align}
Expanding for small $\Delta$ we get 
\begin{equation}
    H_S\left|\psi\right>  = \frac{\hbar^2}{2\tilde{C}_{12}\Delta^{2}}\left|\Psi\right> + \mathcal{O}\left( \partial ^2 \psi\right),
\end{equation}
with $\mathcal{O}\left( \partial ^2 \psi\right)$ indicating additional states which only depend on the second (non-discrete) derivatives of $\psi$. 
Finally, since in the $\Delta^{2} \rightarrow 0$ limit the derivatives thermal average do not depend on $\Delta^{2}$, and using the lemma above, we find
\begin{equation}
    s \leq \exp \left[ -\frac{\beta\hbar^2}{\tilde{C}_{12}\Delta^{2}} + \mathcal{O}\left(1\right)\right].
\end{equation}
Hence, for small $\Delta^{2}$, $s$ becomes exponentially small, meaning that for fine discretization  of the flux the sign problem will be exponentially bad as we decrease the temperature, increase the coupling strength or decrease $\Delta$. The argument can easily be generalized to larger systems.

\section{QMC simulation of superconducting flux circuits: Technical details}
\label{sec:exp_details}
We simulate the flux circuits using the permutation matrix representation (PMR) QMC technique introduced and described in Refs.~\cite{ODE,ODE2,pmr}. The method is based on the off-diagonal expansion of the quantum partition function~\cite{ODE,ODE2} wherein the Hamiltonian is cast in the form, 
%****************************************************************
	\beq \label{eq:basic}
	H=\sum_{j=0}^M D_j P_j  \,,
	\eeq	%****************************************************************
with the matrices $D_j$ and $P_j$ represented in the computational basis, $\{|z\rangle\}$. For $j=0$, $P_j$ is the identity, $P_0=\mathbb{1}$, and for $j>0$, $P_j$ are permutation operators with no fixed points, i.e., $P_{j>0}$ is purely off-diagonal. Conversely, $D_j$ is purely diagonal. This formulation allows us to write the canonical partition function as the sum
%****************************************************************
	\beq \label{eq:z1}
	Z =\sum_{\{z\}} \sum_{q=0}^{\infty}  \sum_{\{S_{{\bf{i}}_q}\}}  D_{(z,S_{{\bf{i}}_q})} \bra{z} S_{{\bf{i}}_q} \ket{z}   \e^{-		\beta [E_{z_0},\ldots,E_{z_q}]} \,,
	\eeq 	%****************************************************************
where $\{S_{{\bf{i}}_q}\}$ is the set of all (unevaluated) products \hbox{$P_{i_q} \ldots P_{i_2} P_{i_1}$} of size $q$,  ${\bf i}_q = (i_1,\ldots,i_q)$ is a multiple index where each individual index $i_j$ (with $j=1\ldots q$) ranges from $1$ to $M$, and the term $e^{-\beta[E_{z_0},\ldots,E_{z_q}]}$ is the divided differences of the exponential function over the multiset of classical energies $[E_{z_0},\ldots E_{z_q}]$~\cite{Whittaker_Robinson,deBoor2005}.
The energies $\left\{E_{z_i}=\langle z_i | H_\D|z_i\rangle \right\}$ are the classical energies of the states $|z_0\rangle, \ldots, |z_q\rangle$, obtained from the action of the ordered $P_j$ operators in the sequence, $S_{{\bf{i}}_q}$, acting on $|z_0\rangle$ through $|z_q\rangle$. We also denote,
%****************************************************************
	\beq
	D_{(z,S_{{\bf{i}}_q})}=\prod_{j=1}^q d^{(i_j)}_{z_j}\,,
	\eeq
%****************************************************************
where $d^{(i_j)}_{z_j}$ is the ``hopping strength'' of $P_{i_j}$ with respect to $|z_j\rangle$, given explicitly by,	%****************************************************************
	\beq \label{eq:hsAppC}
	d^{(i_j)}_{z_j} = \langle z_j | D_{i_j}|z_j\rangle \,. 
	\eeq
%****************************************************************
For further details we refer the reader to Ref.~\cite{pmr}. 

To simulate flux circuit Hamiltonians with PMR we first cast the discretized Hamiltonian, Eq.~\eqref{sec:theory-eq:discrete_H},
in the form of Eq.~(\ref{eq:basic}), by regrouping terms. Specifically, we rewrite the equation as    %****************************************************************
\beq 
	H=\sum_{j=-M}^{M} D_j P_j  \,.
	\eeq
%****************************************************************
For $j=0$, $P_0=\sum_I |I\rangle\langle I|$, which is the identity (as expected), and $D_0 = \sum_I \Big(V_I + \frac{2\mu\hbar^2}{\Delta^2}\Big)|I\rangle\langle I|$ , 
where $\mu = \sum_{k=1}^{n} \mu_k$. For $j\ne0$, the permutation operators are then given by,	%****************************************************************
	\begin{equation}
    P_{\pm j} = \sum_I  |I_j^{(\pm)}\rangle \langle I| \,.
    \end{equation}
%***************************************************************
In matrix form, $P_j$ and $P_{-j}$ are zero everywhere except for the first upper and lower diagonal respectively being unity. These two operators have the property, $P_j P_{-j}=\mathbb{1}$. The corresponding diagonal operators are  $D_j = \sum_I \Big(-\frac{\mu_j\hbar^2}{\Delta^2}\Big)|I\rangle\langle I|$ for all $j$. 

The QMC configuration for this model is given by ${\cal C}=\{|I\rangle, S_{{\bf{i}}_q}\}$---i.e., a basis state $|I\rangle$ and a sequence of permutation operators $S_{{\bf{i}}_q} = P_{i_q} \cdots P_{i_2} P_{i_1}$ that evaluate to the identity operation. The initial configuration ${\cal C}_0$ of the simulation consists of a random basis state $|I_0\rangle$ and the empty sequence $S_{i_0}$. The weight of this initial configuration is thus
%****************************************************************  	
\begin{equation}
	W_{{\cal C}_0} = D_{(I_0, S_0)}  e^{-\beta [E_{I_0}]}=e^{-\beta E_{I_0}},
\end{equation}	%**************************************************************** 
where $E_{I_0}$ is the classical energy corresponding to $|I_0\rangle$.\\

%\noindent
We next review the QMC updates used in our simulations. \\
\noindent {\textit {Classical Move (short-step):}} In this move we shift the basis state by one grid spacing in a randomly chosen direction $|I\rangle \rightarrow |I_k^{(\pm)}\rangle$ for some randomly chosen direction $k$, leaving the permutation operator sequence $S_{{\bf{i}}_q}$ unchanged. If we let ${\cal C}$ be the configuration before the move and ${\cal C'}$ the configuration after the move,  then we can express the acceptance probability for a classical move as
%****************************************************************
	\beq\label{eq:Pmet}
	p = \min \left( 1,\frac{W_{{\cal C}'}}{W_{{\cal C}}} \right)=
	\min \left( 1,\frac{\e^{-\beta [E_{{\cal C}'}]}}{\e^{-\beta [E_{{\cal C}}]}} \right)\,,
	\eeq
%****************************************************************
where $\e^{-\beta [E_{{\cal C}}]}$ is a shorthand for $\e^{-\beta [E_{z_0},E_{z_1},\ldots,E_{z_q}]}$ of configuration ${\cal C}$ (and ${\cal C}'$).\\
\noindent {\textit {Classical Move (long-step):}} For the long-step classical move we follow the same procedure except instead of a step size of one, we use a step size $M$. We take $M$ to be proportional to $D/\Delta$ where $D$ is the distance between minima of the potential wells. This can be done because we know $\textit{a priori}$ that the potential energy surface has a finite number of well separated minima. These larger steps help eliminate the chance of being stuck (for a long time) in a local minimum, resulting in dramatically reduced equilibration time. Again, we take ${\cal C}$ to be the configuration before the move and ${\cal C'}$ to be the configuration after, allowing for the same acceptance probability seen in Eq.~(\ref{eq:Pmet}).\\
\noindent {\textit {Block-swap Move:}} A block swap update involves the change of both the basis state and sequence of permutation operator $S_{{\bf{i}}_q}$. In this move a random position $j$ in the product $S_{{\bf{i}}_q}$ is picked, and the product is split into two (non-empty) sub-sequences, $S_{{\bf{i}}_q}=S_1 S_2$, with $S_1 = P_{i_q} \cdots P_{i_{j+1}}$ and $S_2 = P_{i_{j}} \cdots P_{i_{1}}$.  The state $\ket{I'}$ at position $j$ in the product can be mapped back to the current state, $\ket{I}$, by $\bra{I'} =  \bra{I} S_1 =  \bra{I} P_{i_1} \dots P_{i_{j}} $,
where $\ket{I}$ and $\ket{I'}$ have energies $E_{I}$ and $E_{I'}$, respectively.  This gives a new block-swapped configuration as ${\cal C}'=\{|I'\rangle, S_2 S_1\}$ and a weight that is proportional to $e^{-\beta [E_{{\cal C}'}]}$
where the multiset $E_{{\cal C}'}=E_{{\cal C}} + \{E_{I'}\} - \{E_I\}$. 
Lastly, we can again express the acceptance probability in the same manner as in Eq.~(\ref{eq:Pmet}), using the aforementioned $E_{{\cal C}'}$.\\
\noindent {\textit {Cycle completion Move:}} 
All moves presented thus far have left $q$ unchanged in that the number of elements in the sequence has remained constant. 
Cycle completion moves, on the other hand, change the value of $q$. In this type of move we randomly pick a point, $j \in [0,q]$, in the sequence. This point defines a subsequence of the form $P_0 P_0$ or $P_{i_j} P_{i_{j+1}} $, with $P_0$ and $P_{q+1}$ corresponding to the identity. The new configuration $\mathcal{C}'$ is then defined by replacing the identified subsequence by an equivalent subsequence of length at most two. The two subsequence $P_{i_j} P_{i_k}$ and $P_{i_j}' P_{i_k}'$ are equivalent if $(P_{i_j} P_{i_k}). {(P_{i_j}' P_{i_k}')}^{-1} = \mathbb{1} $. Because we can interpret  $P_0^{-1} = P_0 = P_j P_j^{-1}$ (for any arbitrary index $j$) and so on, the cycle completion move can grow and shrink the sequence. The acceptance probability is the same before (see Eq.~(\ref{eq:Pmet})). \\

The last step in QMC is to take a measurement of some observable. Below we briefly describe the diagonal measurement in our QMC model.\\
\noindent {\textit {Diagonal measurement:}} 
We know that any diagonal operator, $\Lambda$, obeys the eignevalue equation $\Lambda|I\rangle=\lambda_I|I\rangle$, where $\lambda_I$ is the eigenvalue of $\Lambda$ for eigenstate $|I\rangle$. 
We also know that that for any configurationm ${\cal C}=\{|I\rangle,S_{{\bf{i}}_q}\}$, there is a contribution to the thermal average $\langle\Lambda\rangle$ given by $\langle I| \Lambda S_{{\bf{i}}_q} |I\rangle=\lambda_I\langle I| S_{{\bf{i}}_q}|I\rangle$, assuming $\Lambda$ is Hermitian. All of this allows for the simple measurement of $\langle\hat{\Phi}\rangle$, since the basis states, $|I\rangle=|i_1,...,i_n\rangle$, are approximate eigenstates of, $\hat{\Phi}_k$, with eigenvalues $\lambda = i_k\Delta$. This is useful because the quantity of interest is $\hat{I}^z$, which is a weighted sum of $\hat{\Phi}$'s---i.e., $\hat{I}^z$ is also diagonal.

\bibliography{ref2}

\end{document}